\documentclass[aps, prr, reprint, superscriptaddress, longbibliography]{revtex4-2}

\usepackage[colorlinks=true,linkcolor=blue,citecolor=blue,urlcolor=blue]{hyperref}
\usepackage[latin1]{inputenc}
\DeclareUnicodeCharacter{00A0}{ }	
\usepackage[separate-uncertainty=true]{siunitx}
\usepackage{graphicx}
\usepackage{gensymb}
\usepackage{amsmath,bm}
\usepackage{braket}
\usepackage[shortlabels]{enumitem}
\usepackage[normalem]{ulem}
\usepackage[usenames,dvipsnames]{color}
\usepackage{soul}
\usepackage{appendix}

\newcommand{\angstrom}{\mbox{\normalfont\AA}}
\newcommand{\rmi}{\mathrm{i}}

\newcommand{\editor}[2]{%
  \expandafter\newcommand\csname #1note\endcsname[1]{%
    \textcolor{#2}{(\textbf{#1:} ##1)}}%
  \expandafter\newcommand\csname #1\endcsname[1]{%
    \textcolor{#2}{##1}}%
  \expandafter\newcommand\csname #1cancel\endcsname[1]{%
    \textcolor{#2}{\sout{##1}}}%
  \expandafter\newcommand\csname #1change\endcsname[2]{%
    \textcolor{#2}{\sout{##1} ##2}}%
  \newenvironment{#1text}{\color{#2}}{\color{black}}
}

\editor{IT}{red}

\begin{document}
\title{Coherent energy exchange between carriers and phonons in Peierls-distorted bismuth unveiled by broadband XUV pulses}

\author{Romain G\'{e}neaux}%
\email{romain.geneaux@cea.fr}
\affiliation{Department of Chemistry, University of California, Berkeley, 94720, USA}
\affiliation{Universit\'{e} Paris-Saclay, CEA, CNRS, LIDYL, 91191 Gif-sur-Yvette, France}
\author{Iurii Timrov}%
\affiliation{Theory and Simulation of Materials (THEOS), and National Centre for Computational Design and Discovery of Novel Materials (MARVEL), \'{E}cole Polytechnique F\'{e}d\'{e}rale de Lausanne, CH-1015 Lausanne, Switzerland}

\author{Christopher J. Kaplan}%
\affiliation{Department of Chemistry, University of California, Berkeley, 94720, USA}

\author{Andrew D.~Ross}%
\affiliation{Department of Chemistry, University of California, Berkeley, 94720, USA}

\author{Peter M. Kraus}
\altaffiliation[Current address: ]{Advanced Research Center for Nanolithography, Science Park 106, 1098 XG Amsterdam, Netherlands}
\affiliation{Department of Chemistry, University of California, Berkeley, 94720, USA}

\author{Stephen R. Leone}%
\affiliation{Department of Chemistry, University of California, Berkeley, 94720, USA}
\affiliation{Chemical Sciences Division, Lawrence Berkeley National Laboratory, Berkeley, California 94720, USA}
\affiliation{Department of Physics, University of California, Berkeley, 94720, USA}

\begin{abstract}
In Peierls-distorted materials, photoexcitation leads to a strongly coupled transient response between structural and electronic degrees of freedom, always measured independently of each other. Here we use transient reflectivity in the extreme ultraviolet to quantify both responses in photoexcited bismuth in a single measurement. With the help of first-principles calculations based on density-functional theory (DFT) and time-dependent DFT, the real-space atomic motion and the temperature of both electrons and holes as a function of time are captured simultaneously, retrieving an anticorrelation between the $A_{1g}$ phonon dynamics and carrier temperature. The results reveal a coherent, bi-directional energy exchange between carriers and phonons, which is a dynamical counterpart of the static Peierls-Jones distortion, providing first-time validation of previous theoretical predictions.
\end{abstract}

\maketitle

\section{Introduction}

\begin{figure*}[t]
 \includegraphics[width=\linewidth]{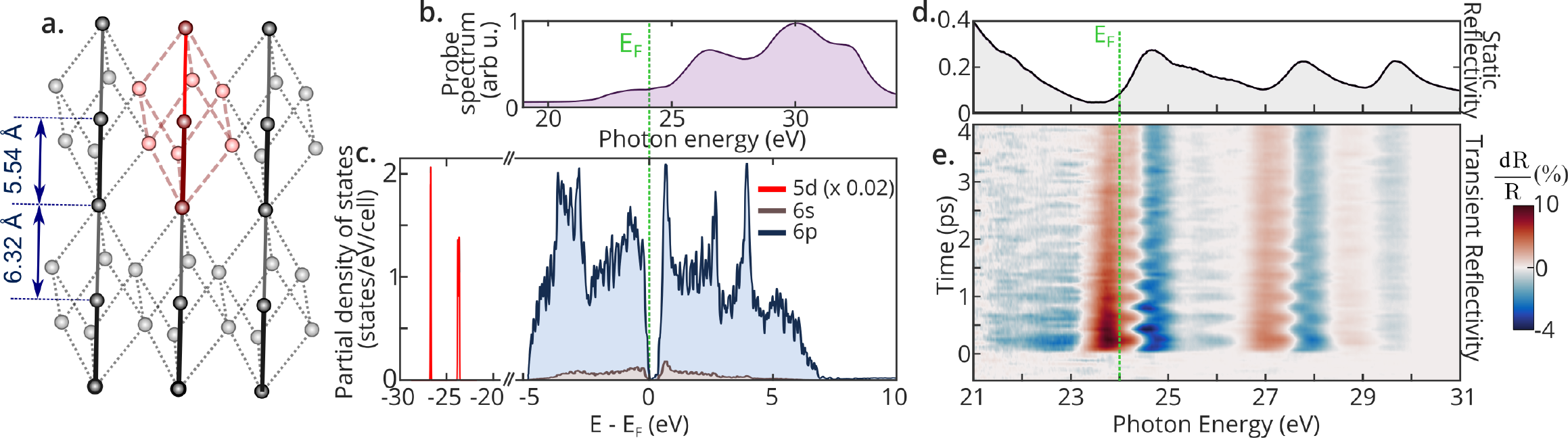}
 \caption{\label{fig1} Properties of bismuth and XUV reflectivity measurements. (a)~Cristalline structure of bismuth, with one rhombohedral unit cell highlighted in red. The Peierls-Jones distortion leads to a dimerized structure, with alternating short and long bonds (thick and thin full lines, respectively). (b)~Measured XUV probe spectrum. (c)~Partial density of states per atom with contributions coming from $5d$, $6s$, and $6p$ electrons, computed using DFT. The $5d$ bands are rigidly shifted by -2.7~eV to match experimental observations (see text). (d)~Static reflectivity of the bismuth sample measured by referencing it to a previously characterized gold mirror. (e)~Transient reflectivity measured over 4~ps with time steps of 60~fs and averaged 30 times. Green dashed lines indicate the Fermi level, which is aligned in all panels following Ref~\cite{Ejiri1981}.}
\end{figure*}

Ultrashort optical pulses can excite materials away from thermal equilibrium, where interactions between the degrees of freedom of solids can be explored. Electron-lattice interactions in particular determine numerous material functionalities: they set fundamental limits on charge-carrier mobilities in photovoltaic applications~\cite{Yu1999}, lead to the formation of electron pairs responsible for superconductivity~\cite{Bardeen1957}, and drive the appearance of charge-density waves~\cite{Gruner1988}. The semimetal bismuth is a particularly important system for which electron-phonon interactions are being explored. A key property of bismuth is its equilibrium crystal structure: the two atoms of its primitive unit cell of the A7 structure~\cite{Chang1986} are displaced towards each other compared to the simple cubic structure. This creates a dimerized structure along the [111] direction (Fig.~\ref{fig1}a) in which the Bi-Bi bond length is alternatively short and long - akin to the one-dimensional distortion described in the seminal work of Peierls~\cite{Peierls1955}. While the structural distortion by itself has a detrimental energy cost, it allows to lift the degeneracy of valence electronic states. This results in electrons occupying states of lower energy, which more than counterbalances the cost of the structural distortion. This phenomenon, known as a Peierls-Jones distortion~\cite{Jones1934,Mott1936}, is analoguous to a Jahn-Teller effect for extended systems. 

Photoexcitation gives the opportunity to tip the delicate energy balance of the system: when electrons are optically promoted from the valence states, less energy is now gained by the lifting of degeneracy. This makes the structural distortion less favorable, with the quasi-equilibrium atomic position shifting towards a more symmetric and metallic structure. Time-resolved techniques allow to watch the system evolve towards this new equilibrium, uncovering spectacularly large and fast atomic motion~\cite{Sciaini2009}. During this relaxation, there is a bi-directional energy exchange between electronic and structural degrees of freedom, which, if it can be captured, grants a unique view into the electron-phonon interactions at the heart of a Peierls-Jones distortion. The light-driven electronic properties have been thoroughly studied using optical pump-probe techniques~\cite{Zeiger1992,Hase2002,Boschetto2008} as well as angle-resolved photoemission~\cite{Papalazarou2012}, while the transient atomic motion has been directly measured using ultrafast x-ray diffraction~\cite{Johnson2008,Fritz2007}. However, probing the electron, hole, and phonon responses simultaneously in a single experiment is out of reach for these approaches. Since the electronic and structural dynamics are inherently strongly coupled, such a capability is highly desirable to obtain a complete and unified picture of the photoinduced dynamics.

In the investigation presented here, photoexcited bismuth is probed using broadband transient reflectivity in the extreme ultraviolet (XUV) spectral range at the Bi $\text{O}_\text{4,5}$ edge. With the help of first-principles simulations based on density-functional theory (DFT)~\cite{Hohenberg:1964, Kohn:1965} and time-dependent DFT (TDDFT)~\cite{Runge:1984, Gross:1985}, we measure both lattice motion and carrier dynamics, garnering separate information on electron and hole temperatures. These carrier temperatures are found to have distinct dynamics and, surprisingly, to oscillate in anti-phase with the oscillations of atomic positions - an hitherto unobserved behavior. We explain this phenomenon by entropy conservation arguments that can be seen as a dynamic equivalent of the static Peierls-Jones distortion. Our results highlight the coherent bi-directional energy exchange between electron and phonon sub-systems on the femtosecond timescale, and provide a compelling illustration of the use of time-dependent measurements to gain insight into time-independent interactions. 
 
\section{Overview of reflectivity measurements, XUV transitions, and observations}

The overview of the basic experiment is outlined in Fig.~\ref{fig1}. The partial density of states (PDOS) of Bi is computed using DFT (see Appendix~\ref{sec:Technical_details_of_calculations}), shown in Fig.~\ref{fig1}c. About 24 eV below the Fermi level ($\text{E}_\text{F}$) lie two groups of bands with $5d$ character, which are split by about 3~eV due to strong spin-orbit (SOC) coupling. The topmost valence bands and lowest conduction bands mainly have $6p$ character. %and are also affected by strong SOC, which yields a large splitting of bands (e.g. the splitting is 1.5~eV at the $\Gamma$ point). 
XUV radiation of appropriate energy can photoexcite the $5d$ semicore electrons to the Fermi level and the lowest empty $6p$ bands, a process that corresponds to the transition at the Bi $\text{O}_\text{4,5}$ edge.

In the experiment, a single-crystal of bismuth with (0001) orientation is probed by XUV radiation created by high-harmonic generation. Using sub-5~fs near-infrared (NIR) driving laser pulses, a xenon high-harmonic target gas and subsequent spectral selection (see Appendix~{\ref{app:Expt_apparatus}}), a continuous spectrum spanning $20-36$~eV is obtained (Fig.~\ref{fig1}b). The static reflectivity of the sample, probed at 66\degree{ }from its surface normal (Fig.~\ref{fig1}d), shows a maximum at 0.6~eV above the Fermi level (which lies at $\text{E}_\text{F} =24$~eV~\cite{Ejiri1981}) corresponding to the maximum in the conduction-band density of states (DOS) as obtained from DFT (Fig.~\ref{fig1}c) and in agreement with earlier work~\cite{Ejiri1981}. The next large peak $3.10\pm0.01$~eV higher in energy corresponds to transitions from the deeper spin-orbit split $5d$ bands to the empty states. A detailed assignment of features is presented together with the band structure in Appendix~\ref{app:Origin_of_peaks}. 

The sample is photoexcited in pump-probe measurements by sub-5~fs NIR pump pulses at an intensity of 0.5-1.5~\si{mJ/cm^2}, using the same setup as in Refs.~\cite{Kaplan2018,Geneaux2020} The pump-induced changes in XUV reflectivity, measured by scanning the pump-probe delay, processed using an edge-referencing method to reduce correlated source noise (see Appendix~\ref{app:Edge_referencing} and Ref.~\cite{Geneaux2021}), are shown in Fig.~\ref{fig1}e. The transient reflectivity shows picosecond-long signals both above and below the Fermi level, among which the most notable features are oscillations of the XUV reflectivity in several energy ranges, but predominantly close to $\text{E}_\text{F}$. These oscillations are chirped, with instantaneous frequencies increasing from 2.4~THz to 2.8~THz as the time-delay gets longer (see Appendix~\ref{app:tw_analysis} for a time-frequency analysis). This is a clear signature of the $A_{1g}$ coherent phonon mode (2.93~THz at equilibrium~\cite{Hase1998}) associated with the Peierls-Jones distortion, whose softening under carrier excitation is well known~\cite{DeCamp2001,Fritz2007}. 
While oscillations of XUV reflectivity have already been observed and attributed to coherent phonon motion~\cite{Papalazarou2008, Weisshaupt2017a,Kato2020}, a detailed understanding of their origin in terms of interactions between excited carriers (electrons and holes) and the lattice (phonons) is lacking. The oscillations could stem from energy shifts of valence bands (due to the coupling of valence electrons to coherent phonons via electron-phonon interactions~\cite{Papalazarou2012}) but also from core-level shifts, which can be particularly sensitive to bond lengths~\cite{Cushing2017,Geneaux2019,Obara2017}, or even from modulations of either hole or electron temperatures around the Fermi level~\cite{Giret2011}. To elucidate their origin, we perform~\textit{ab initio} calculations based on TDDFT with the aim of understanding the time-dependent (dynamical) effects of the coherent $A_{1g}$ displacement on the XUV spectrum of the material. 

\section{First-principles calculations of XUV absorption}

The bismuth $\text{O}_\text{4,5}$ absorption spectrum is first calculated at equilibrium using TDDFT (see Appendix~\ref{sec:Technical_details_of_calculations}). At the incidence angle used in the experiment, the XUV penetration depth is 13.6~nm at 24~eV (the Fermi level energy). This is much larger than the extent of the surface states of Bi which have two-dimensional character \cite{Ast2001}. Thus, in what follows we focus solely on the bulk electronic and phonon properties of bismuth.

\begin{figure}[t]
 \includegraphics[width=\linewidth]{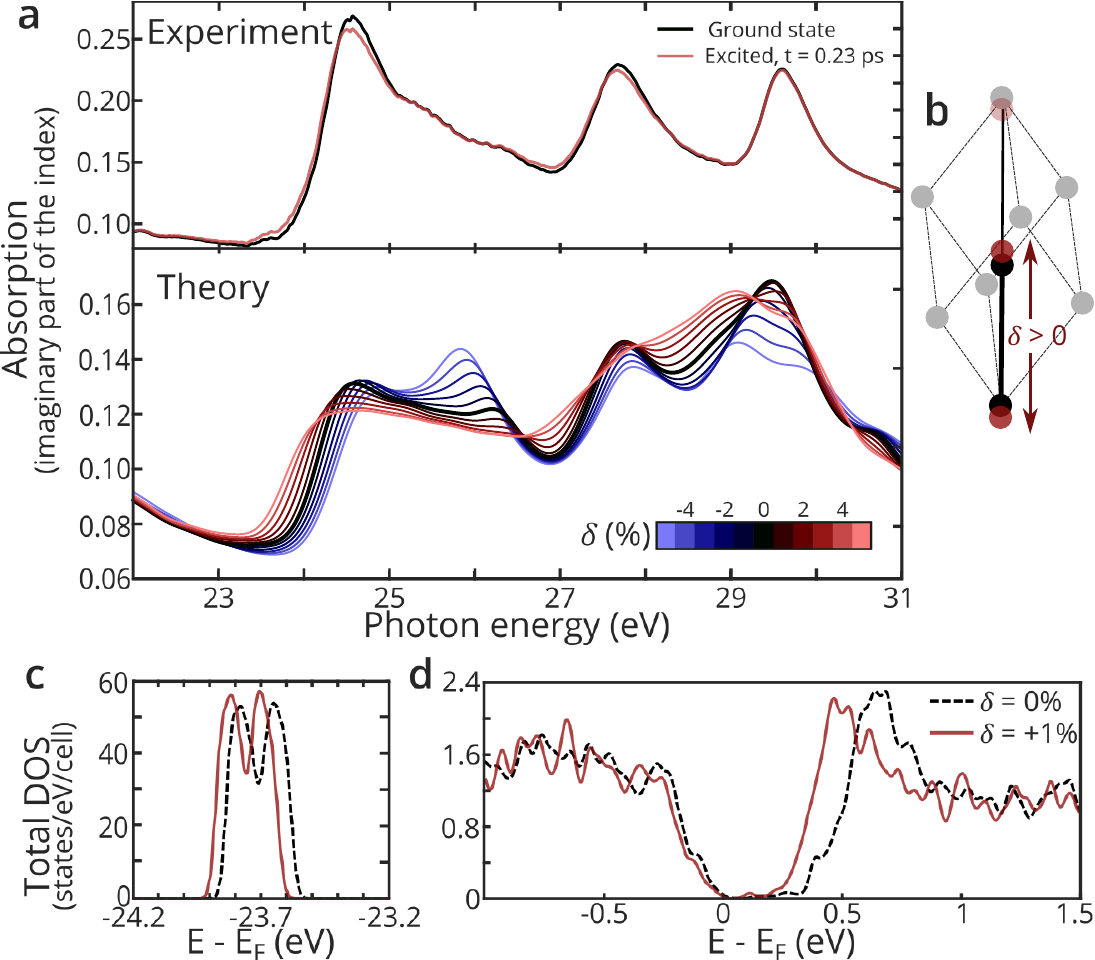}
 \caption{\label{fig2} Influence of atomic displacements on the XUV absorption spectrum. (a)~Experimental and calculated absorption spectra, expressed as the imaginary part of the refractive index. The experimental spectrum, obtained via Kramers-Kronig transformation, is shown at equilibrium (black) and at $t=0.23$ ps (red). The computed absorption spectrum is shown with $\delta$ varying from -5\% (blue) to +5\% (red). (b)~Illustration of the Bi primitive unit cell and definition of $\delta$. (c)~Total DOS for bismuth in the semicore region and (d)~near the Fermi level. These are shown at equilibrium (black dashed line) and with a 1\% increase in the quasi-equilibrium inter-atomic distance (red solid line).}
\end{figure}

The theoretical equilibrium distance between nearest Bi atoms along the trigonal axis is 5.50~\angstrom, which is in very good agreement with the experimental value at room temperature of 5.54~\AA~\cite{Murray2005,Fritz2007}. To allow for comparison, the experimental absorption spectrum is obtained by inverting the measured reflectivity of Fig.~\ref{fig1}d using a Kramers-Kronig procedure previously described~\cite{Geneaux2020}. Fig.~\ref{fig2}a displays the experimental and theoretical absorption spectra, with the latter being rigidly shifted down by 2.7~eV to match the experimental position of the absorption edge. Such a shift is necessary because of the overdelocalization of the $d$ electronic states when using (semi-)local exchange-correlation (xc) functionals in (TD)DFT, which are responsible for self-interaction errors~\cite{Perdew:1981} and lead to the imprecise position and shape of such states.
Apart from this shift, the overall shape of the computed spectrum is in good agreement with the experimental one, with discrepancies only in the relative intensity of the above-edge peaks, which correspond to the various conduction band extrema~\cite{Ejiri1981}. It is worth noting that more accurate absorption spectra can be obtained by solving a Bethe-Salpeter equation on top of a $GW$ calculation \cite{Strinati1988, Rohlfing2000, Onida2002, Palummo2004}; however, for the purposes of this work the absorption spectrum computed at the TDDFT level is fully sufficient.
%are fingerprints of the conduction bands dispersion in Bi
%We note that more advanced simulations based on many-body perturbation theory~\cite{Onida:2002} could refine and improve the agreement with the experimental absorption spectrum, but such treatment is beyond the present study; as we will see, this does not prevent the extraction of quantitative trends.

The calculation is now repeated while changing the Bi-Bi distance in the unit cell along the trigonal axis, which is denoted by $\delta$ and is expressed in percent of the initial Bi-Bi distance; positive $\delta$ corresponds to a repulsion of Bi atoms away from each other, as illustrated in Fig.~\ref{fig2}b. This approach allows us to observe changes in the imaginary part of the refractive index of Bi due to the coupling of photoexcited carriers with the lattice vibrations. The computed absorption spectrum is shown in Fig.~\ref{fig2}a for $\delta$ evolving from -5\% to 5\%. The computation shows that the displacement along the phonon coordinate leads to a blueshift (resp. redshift) of the edge for a contraction (resp. elongation). For small displacements, the shift is linear with $\delta$ at a rate of 9.7 meV/pm. This value can be seen as the local gradient of the potential energy surface of the semicore-excited state along the $A_{1g}$ phonon coordinate. Several eVs above the absorption edge, the behavior becomes more complicated. This is because XUV absorption probes the entire Brillouin zone and thus the contributions of higher conduction bands at various momenta are summed together. Since the electron-phonon coupling varies strongly with momentum \cite{Papalazarou2012}, this results in a more complicated trend than a rigid shift. Additionally, absorption from the $5d_{3/2}$ core-level further complicates the spectrum above 27~eV (see Appendix~\ref{app:Origin_of_peaks}). We thus focus on the shift near the absorption edge. We can elucidate the origin of this shift by looking at the DOS, shown for a repulsion of $\delta = +1\%$  in Figs.~\ref{fig2}c-d. As is customary in DFT, the computed DOS at and out of equilibrium must be arbitrarily aligned to some energy; we choose here to align them at the Fermi energy. This is contrary to previous calculations interpreting photoemission experiments in bismuth where the electronic bands were aligned at the $5d$ level energy~\cite{Papalazarou2012}, as the removal of an electron from the system during photoemission does not keep the Fermi energy constant. The conduction band (CB) states shift towards the Fermi level by about 100~meV while the binding energy of the $5d$ semicore states increases by approximately 45~meV. This is consistent with the fact that Bi in the excited state is more symmetric and metallic~\cite{Shick:1999, Wu:2019}: the indirect negative gap is closing, while the Coulombic repulsion between ionic cores is reduced, increasing the binding energy of the $5d$ levels. Here, the CB shifts more than the $5d$ levels, resulting in a decrease of the transition energy; however, it is important to note that this will not necessarily be true for all systems. Finally, we briefly explored the impact of the A$_{1g}$ phonon mode on the electronic structure near the L~point, where doping-induced topological phase transitions were observed \cite{Jin2020}. As shown in Appendix~\ref{app:lpoint}, we do not find evidence of phonon-induced phase transition, but it must be stressed that our level of theory is not well suited to describing such fine features at low energies.

\section{Decomposing the transient reflectivity trace: carrier and phonon dynamics}

\begin{figure}[b]
 \includegraphics[width=\linewidth]{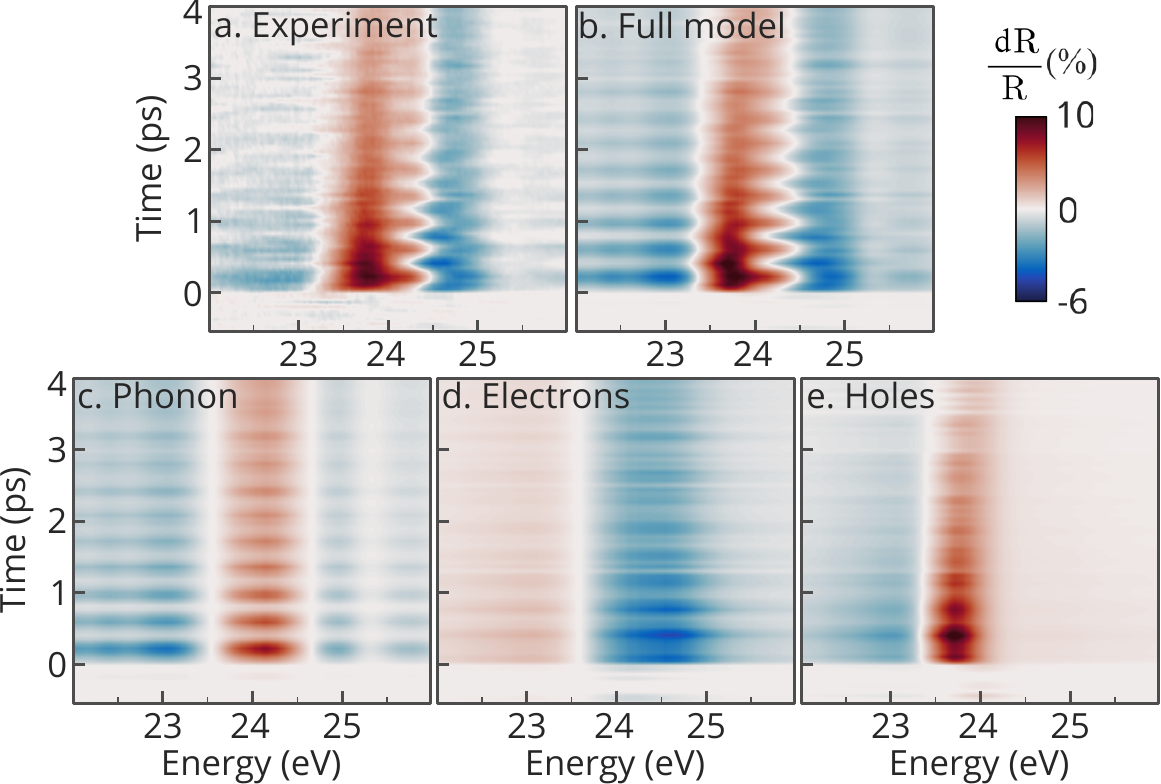}
 \caption{\label{fig3} Phonons and carriers contributions to transient reflectivity. Transient experimental (a) and modeled (b reflectivity spectra. Individual phonon (c), electron (d), and hole (e) contributions are extracted from the fit.}
\end{figure}

Now that we linked the A$_{1g}$ phonon motion with a shift of the absorption edge, it is possible to extract quantitative information from the XUV transient reflectivity measurement. The procedure follows established methods successfully applied to elementary semiconductors~\cite{Kaplan2018}, transition metal dichalcogenides~\cite{Attar2020}, and perovskites~\cite{Verkamp2019}. We decompose the measured dynamics into three contributions: phonon motion (edge shift), holes photoexcited in the valence band and electrons in the conduction band.
The decomposition attempts to recover three variables at each time-delay: the electronic and hole temperatures, $T_{elec}(t)$ and $T_{holes}(t)$, as well as the phonon-induced spectral shift $\Delta E(t)$, as explained in Appendix~\ref{app:Decomposition}. The combination of all contributions is fitted to the experimental transient reflectivity by a least-square minimization procedure in the interval $23.0 - 25.5$~eV. The experimental and converged result for the transient reflectivity are shown in Figs.~\ref{fig3}a and b, and the model is separated into each of its components in Figs.~\ref{fig3}c--e.
 The model is able to successfully reproduce the experimental features. The individual contributions each show distinct behaviors and spectral signatures, which are separable due to the broadband nature of the spectroscopic observable. We remark that due to the reflection geometry, the total transient reflectivity is not given by the simple sum of each component, and that the electrons and holes contribute signals away from their respective bands~\cite{Kaplan2018}. While the ability of the model to extract accurate absolute carrier temperatures deserves further exploration, a wealth of information on the separate dynamics of holes, electrons, and phonons can be extracted.
 
 %In turn, the time evolution of the three components offers unique insight\IT{s} in the coupled \ITcancel{electronic and structural} dynamics \IT{of holes, electrons, and phonons}. \ITnote{TODO: Add some more discussion about nice results which can be possibly extracted from Fig.~\ref{fig3}. Or at least comment what do we see there (e.g. holes oscillate less than electrons - why?, electrons oscillate in the antiphase with phonons, etc.)}
 %
 \begin{figure}[t]
 \includegraphics[width=\linewidth]{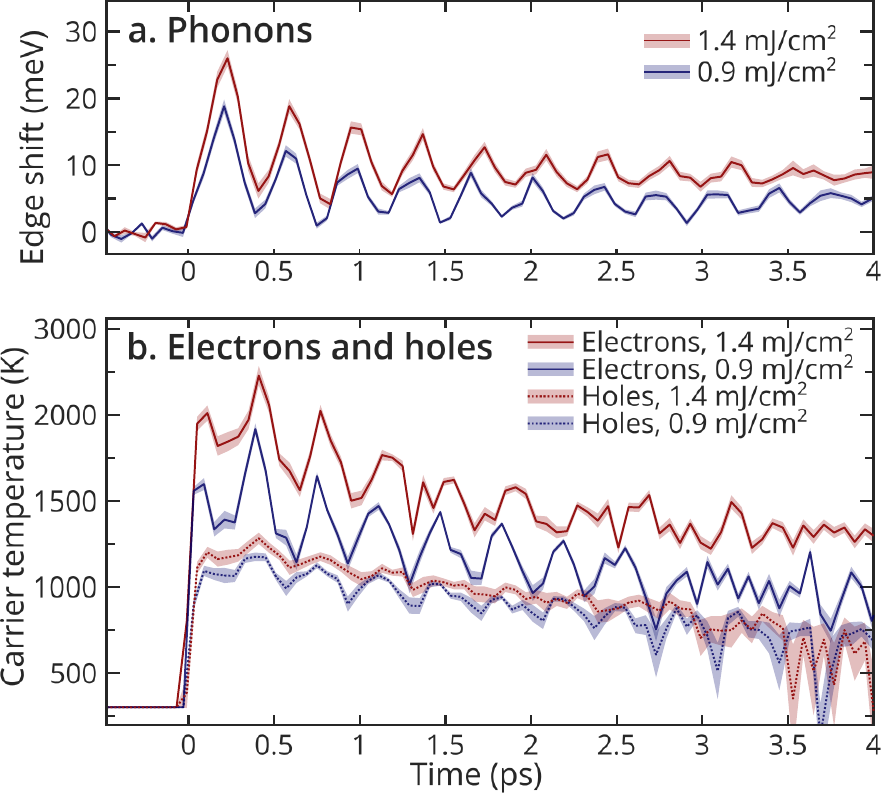}
 \caption{\label{fig4} Dynamics of edge shift and carrier temperatures. (a)~Edge shifts and (b) carrier temperatures extracted from the model used to fit the experimental data, at pump fluences of 0.9~\si{mJ/cm^2} (blue) and 1.4 \si{mJ/cm^2} (red). In (b), electrons and holes are in full and dashed lines, respectively. Shaded areas represent the 95\% confidence interval of the fit, obtained from the Jacobian of the minimized objective function.}
\end{figure}
Figure~\ref{fig4} shows the extracted variables (edge shift and two carrier temperatures) for two sets of measurements taken at different pump fluences. The phonon-induced shift (Fig.~\ref{fig4}a) oscillates with a very visible phonon-softening and fluence-dependent motion amplitude. The shift can be converted to real space motion using the gradient determined by TDDFT of 9.7~meV/pm: we obtain maximum excursions of $1.9\pm 0.1$~pm and $2.6\pm 0.1$~pm for 0.9~\si{mJ/cm^2} and 1.4~\si{mJ/cm^2} fluences, respectively. This conforms well with the values measured using x-ray diffraction~\cite{Fritz2007} of $1.7$~pm and $2.9$~pm at 1.2~\si{mJ/cm^2} and 1.7~\si{mJ/cm^2}. Furthermore, in Appendix~\ref{app:Atomic_motion} we directly juxtapose the extracted time-dependent atomic positions with previous \textit{ab initio} calculations~\cite{Giret2011}, for a variety of pump fluences. The comparison (Fig.~\ref{figCompGiret}) indicates a striking agreement in both the magnitude of actual atomic displacement and the frequency of the motion, for both fluences used in the experiment. The results and comparisons strongly supports that the all-optical XUV technique yields accurate and quantitative measurements.
%in a strikingly similar manner as the atomic motions measured by x-ray diffraction~\cite{Fritz2007} or obtained by \textit{ab initio} calculations~\cite{Giret2011},  The shift can be converted to real space motion using the ratio previously determined by TDDFT ratio of 9.7 meV/pm: we obtain maximum excursions of $1.9\pm 0.1$ and $2.6\pm 0.1$ pm for 0.9 and 1.4 \si{mJ/cm^2}, respectively. This can be compared to the diffraction results of Fritz et al.~\cite{Fritz2007} of $1.7$ and $2.9$ pm at 1.2 and 1.7 \si{mJ/cm^2}. Therefore we see that the XUV experiment provides a reasonably quantitative measurement of the real-space atomic motion. 

\section{Discussion}
In addition to the atomic motion, the measurements contain another piece of the puzzle -- the dynamics of both electrons and holes. We stress that the accuracy of the method in recovering meaningful absolute values for electron and hole temperatures deserves further investigation before being interpreted, as was recently done for nickel \cite{Chang2021}. However, the relative evolution of the carrier temperatures is insightful; as seen in Fig.~\ref{fig4}b, they decay with markedly different time constants of $\tau_{elec} = 1.5\pm 0.4 $ ps and $\tau_{holes} = 3.8\pm 1.2$~ps. This suggests that each carrier couples differently to the coherent phonon motion, which might be explained by their widely different optical masses~\cite{Timrov:2012}. Both temperatures show a delayed maximum at $\sim 0.3$~ps which is consistent with the carrier thermalization time~\cite{Timrov:2012}. Most intriguingly, the experiment captures pronounced oscillations of both temperatures, at frequencies closely following the atomic motion. However, these oscillations are out of phase: when the atoms are driven away from their equilibrium position (edge shift is maximum, largest elongation), the carrier temperature diminishes, and vice-versa. The decay time of the oscillations is slightly longer for lower pump fluences, following the coherent phonon behavior \cite{Giret2011}.

The excitation of coherent phonons is well described by the displacive excitation mechanism~\cite{Zeiger1992a}, which postulates that the quasiequilibrium $A_{1g}$ nuclear coordinate depends linearly on electronic temperature \footnote{Note that the displacive excitation mechanism of coherent phonons as described by Zeiger et al. \cite{Zeiger1992a} can be formulated equivalently with the nuclear coordinate depending linearly on either the electronic temperature or the carrier density}. This leads to damped oscillations of atomic position, as observed here and in many other experiments. Importantly, an angle-resolved photoemission study \cite{Greif} managed to measure both photoelectron  diffraction and emission  signals in the same experiment, gathering  simultaneous electronic and structural information. This way, the relative phase between phonon displacement and electronic shifts was found to match the DECP process \cite{Hase1998}. However, the observation here is fundamentally different: we probe the relative phase between phonon displacement and electronic temperature. In fact, the DECP mechanism does not explain why the electronic temperature itself oscillates. Instead, we relate to a reported thermodynamical model that anticipated this result~\cite{Giret2011}, which our experiment confirms for the first time. This model succeeded in reproducing x-ray diffraction measurements of bismuth~\cite{Fritz2007} in a particularly remarkable way, since it only has two adjustable parameters. In addition, it predicted antiphase oscillations of the electronic temperature and rationalized them as the way by which the system counteracts the increased entropy created by atomic displacements.

\begin{figure}[t]
 \includegraphics[width=\linewidth]{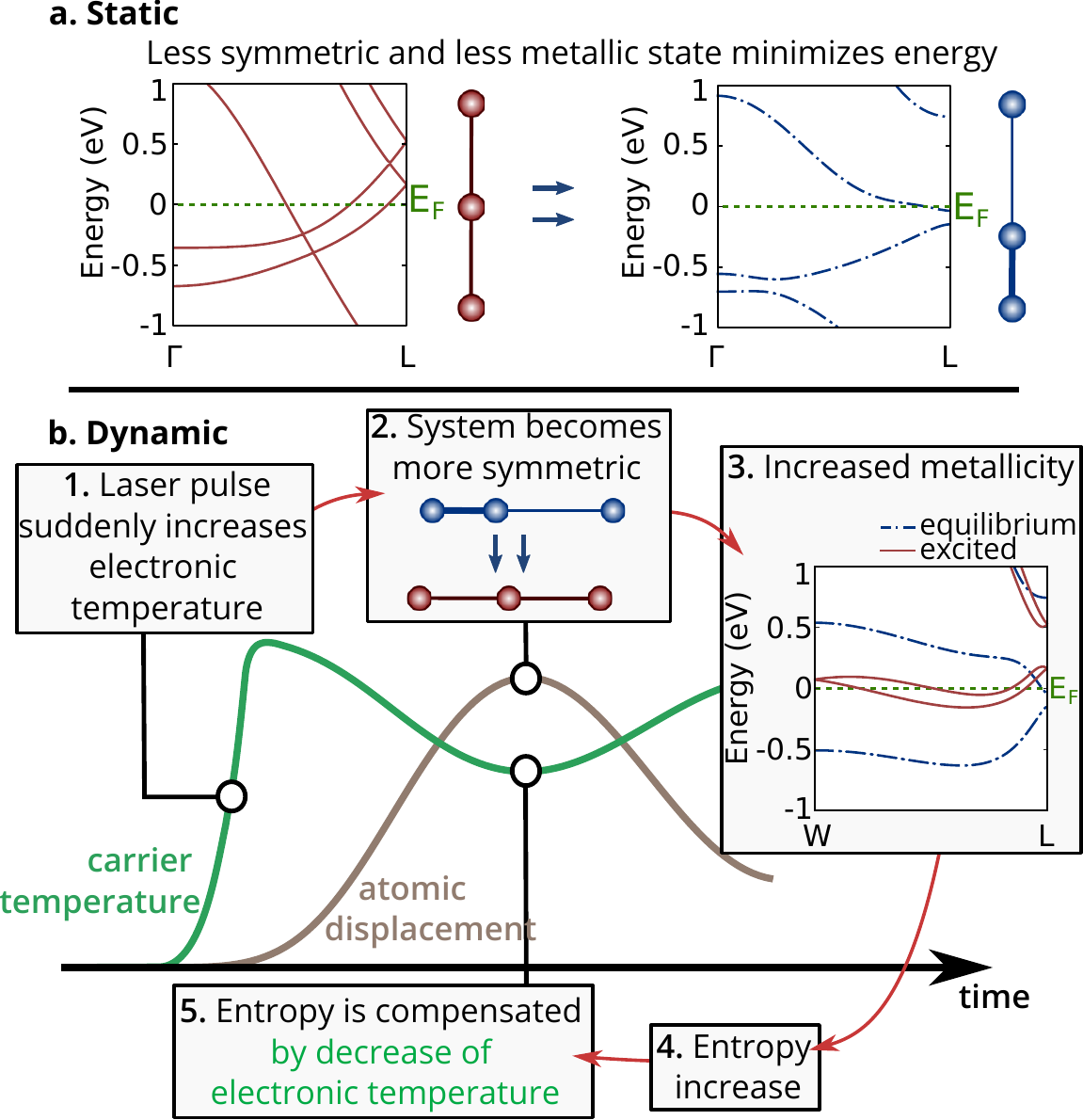}
 \caption{\label{fig5} Static and dynamic energy balance by the Peierls-Jones mechanism. (a) The static structure of bismuth is governed by the Peierls-Jones distortion: the total energy is minimized by going to a distorted structure, whose band structure along the $\Gamma$-L direction as calculated by DFT is shown. (b) In the time domain, the laser pulse suddenly changes the electronic temperature (green curve), triggering an oscillations of atomic positions (gray curve). The band structure (calculated by DFT) is shown along W-L for the equilibrium (dash-dotted blue line) and excited geometries (solid red line). It shows that the system is more metallic for large displacements, increasing the electronic entropy. This excess of electronic entropy is compensated by a concomitant decrease of electronic temperature.}
\end{figure}

Consider the \textit{static} Peierls-Jones distortion, illustrated in Fig.~\ref{fig5}a: at equilibrium, the total energy of the system (sum of an ionic and an electronic part, respectively repulsive and attractive) is reduced by going to a less metallic and less symmetric state. The \textit{dynamic} behavior of the system seems to be similar, with phonons and temperatures oscillating in quadrature as the system attempts to balance structural and electronic energies. Understanding the basis of the thermodynamic model of Giret et al.~\cite{Giret2011}, we propose the mechanism sketched in Fig.~\ref{fig5}b: first, the laser pulse suddenly deposits energy into the system, which increases the carrier temperatures and suddenly changes the quasi-equilibrium position $u_{eq}$ along the phonon coordinate. The total energy thereby deposited is slowly relaxed by energy transfer to the lattice on time scales longer than 10~ps~\cite{Fritz2007,Giret2011}. Thus, on the femtosecond timescale studied here, the total energy, the entropy and $u_{eq}$ are all almost constant after laser excitation. The new value of $u_{eq}$ set by photoexcitation then drives oscillations of atoms around the new position. When the atomic displacement $u$ is maximum, the system is temporarily more symmetric, and in turn, more metallic (see band structures in Fig.~\ref{fig5}b), which increases the electronic entropy~\cite{Giret2011}. Because the entropy must remain almost constant, this excess has to be balanced, and the experiment shows that this is realized by a diminution of the carrier temperature. The observation of this bi-directional energy exchange is possible only because both carrier and structural degrees of freedom are probed simultaneously. In turn, the dynamic behavior of the system can be seen as a time-dependent counterpart of the static Peierls-Jones distortion. 

In summary, we investigated the photoexcited response of the semimetal bismuth using transient broadband XUV spectroscopy and \textit{ab initio} calculations. In a single experiment, real-space motion and carrier dynamics are quantified. We find that the atomic displacement and carrier temperatures oscillate in quadrature, which is a direct view into the energy balance of this Peierls-distorted system and confirms previous theoretical predictions~\cite{Giret2011}. Our approach can be applied to similar systems, with the limitation that materials showing valence or core-excitonic effects would require more advanced computations based on many-body perturbation theory. Examples of materials hosting Peierls distortions are the other group~V elements (As, Sb, black phosphorus), known to also show dimerized structures \cite{Gaspard2016}. Likewise, the structures of Te, Se, and many compounds they form with group~V elements can be seen as n-merizations ($n>2$) and might show comparable electron-phonon dynamics \cite{Shu1988, Imamov1971845}. For instance, a similar behavior was recently observed in the 2D charge density wave material $\text{TaSe}_2$~\cite{Zhang2020} using angle-resolved photoemission. In comparison, the XUV transient technique is bulk-sensitive, has 5-fs time resolution and can probe all bands with allowed orbital character. As such, this work offers a complementary perspective and is particularly well suited to future studies of the highly coupled electron-phonon dynamics accompanying ultrafast light-induced phase transitions~\cite{Gerber:2017}. More generally, the results provide an illustration of how time-resolved studies capturing the dynamics of several degrees of freedom are able to yield insight into fundamental, time-independent interactions between them.

Computational and experimental data that support the findings of this study are available on the Materials Cloud Archive \cite{MaterialsCloudArchive2021}.

\section*{Acknowledgements}
We thank H.-T. Chang for helpful discussions on the analysis and fitting of transient XUV data. Investigations at UC Berkeley were supported by the Defense Advanced Research Projects Agency PULSE Program Grant W31P4Q-13-1-0017 (concluded), the U.S. Air Force Office of Scientific Research Nos. FA9550-19-1-0314, FA9550-20-1-0334, FA9550-15-0037 (concluded), and FA9550-14-1-0154 (concluded), the Army Research Office No. WN911NF-14-1-0383, and the W.M. Keck Foundation award No. 046300-002. I.T. acknowledges funding from the Swiss National Science Foundation (SNSF), through Grant No. 200021-179138, and its National Centre of Competence in Research (NCCR) MARVEL. Computer time was provided by CSCS (Piz Daint) through Project No. s836. R.G. acknowledges support from ``Investissements d'Avenir'' LabEx PALM (ANR-10-LABX-0039-PALM).

%%%%%%%%%%%%%%%%%%%%%%%%%%%%%%%%%%%%%%%%%%%%%%%%%%%%%%%%%%%%
%%%%%%%%%%%% Appendices %%%%%%%%%%%%%%%%%%%%%%%%%%%%%%%%%%%%
%%%%%%%%%%%%%%%%%%%%%%%%%%%%%%%%%%%%%%%%%%%%%%%%%%%%%%%%%%%%

\appendix

\section{Experimental methods}
\label{app:Expt_methods}

\subsection{Experimental apparatus}
\label{app:Expt_apparatus}

The experimental apparatus uses a carrier-envelope phase stable titanium-sapphire laser system (Rainbow CEP4 oscillator + Femtopower amplifier from Femtolasers, pumped by a DM30 laser from Photonics Industries), which delivers 1.8 mJ, 25 fs pulses at 1kHz. They are broadened in a neon-filled hollow-core fiber (1.5 bar static pressure). A set of 12 double-angle chirped mirrors (PC70, Ultrafast Innovations) and a 2~mm-thick ADP crystal control the spectral phase of the pulse, yielding a duration of about 4~fs. Further details can be found in Refs.~\cite{Kaplan2018,Geneaux2020}. 

A 60:40 beamsplitter separates the incoming pulse. The most intense part (480 \si{\micro J}) is focused by a 45~cm spherical mirror into a xenon-filled gas cell ($\sim 20$ Torr) to produce broadband high-harmonic radiation. A 50~nm titanium filter removes the generating infrared light and cuts XUV radiation above 40~eV. This is important for the spectrometer grating to function properly: it prevents second order diffraction of high energy components to overlap with the Bi edge region where the experiment is performed. The XUV is then focused onto the sample by a toroidal mirror. The remaining 40\% of the initial short infrared pulses are recombined collinearly with the XUV light using a hole mirror and used as a pump. The pump and probe beams (p- and s-polarized, respectively) impinge on the sample surface with a 66\degree{ }angle from the sample normal. The pump light is then removed by an 150~nm-thick aluminum filter, and the spectrum of the reflected probe beam is measured using an aberration-corrected diffraction grating (Shimadzu 30-006) and an XUV CCD camera (Princeton Instruments). The spectral resolution at 25 eV is $\sigma = 13$ meV as determined using atomic absorption lines. 

The sample is a single-crystal of bismuth with (0001) orientation, polished with roughness \textless10~nm (purchased from Princeton Scientic).

\subsection{Edge-referencing of transient reflectivity data}
\label{app:Edge_referencing}

\begin{figure}[h!]
 \includegraphics[width=\linewidth]{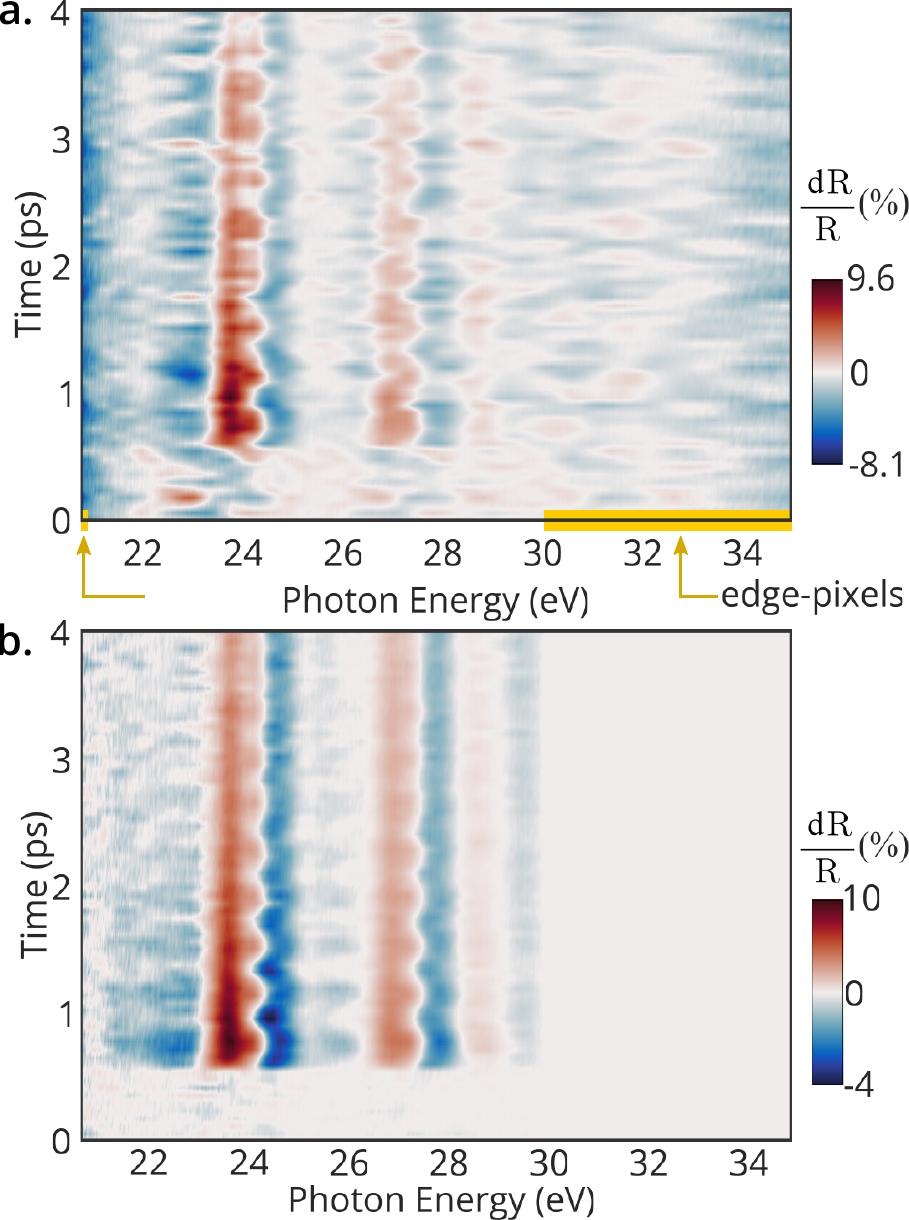}
 \caption{\label{figSM1} Edge-referencing. (a) Measured transient reflectivity trace. Yellow shaded areas indicate the regions chosen as edge-pixels for the noise reduction procedure. (b) Filtered transient reflectivity trace.}
\end{figure}

At each time delay between pump and probe, we measure two spectra $I_\text{on}$ and $I_\text{off}$, with the pump on and off, respectively. The transient reflectivity is then defined as $\frac{I_\text{on}-I_\text{off}}{I_\text{off}}$. The dominant noise source arises due to fluctuations of the XUV flux. They are mitigated using an edge-referencing procedure that is described in detail in Ref~\cite{Geneaux2021}. The edge-pixels were taken as regions without transient signal; namely, photon energies below 21 eV and between 30 and 35 eV. We verified that a slight change of these regions did not influence the final transient reflectivity. Figure~\ref{figSM1} presents the original and filtered transient reflectivity, showing a significant enhancement of the signal-to-noise ratio.

\section{Time-frequency analysis of chirped reflectivity oscillations}
\label{app:tw_analysis}

\begin{figure}[h!]
 \includegraphics[width=\linewidth]{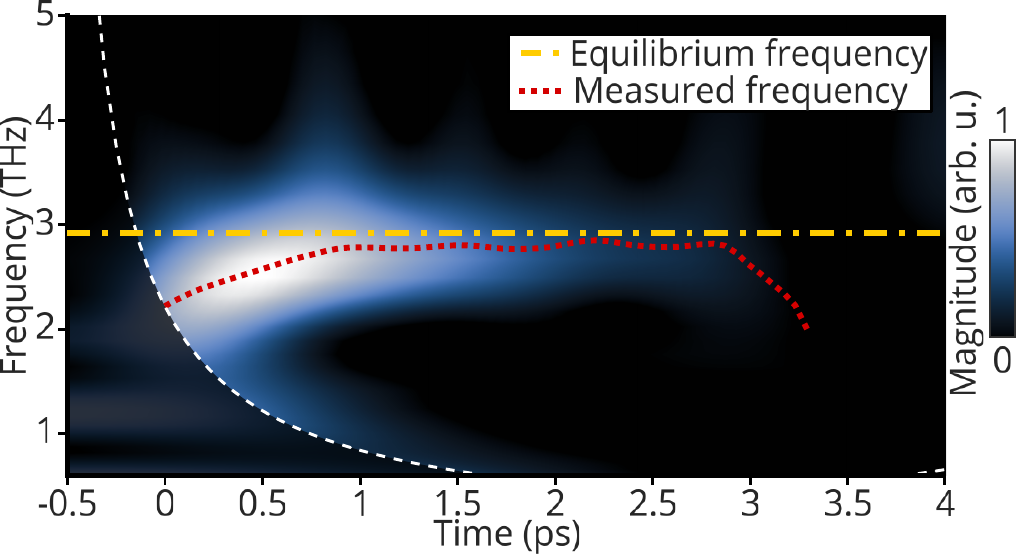}
 \caption{\label{figGabor} Time-frequency analysis. Continuous wavelet transform of the transient reflectivity at $24\pm0.5$~eV using the Morlet wavelet. The instantaneous frequency (dashed red line) evolves towards the equilibrium frequency (dash-dotted yellow line) as time-delay increases. The dashed white line indicates the cone of influence of the CWT, which is the region of the wavelet spectrum in which edge effects become important.}
\end{figure}

Under intense photoexcitation, bismuth is known to undergo bond softening \cite{Hase2002a}, which makes the oscillation frequency depend on the motion amplitude. This was shown to be due to carrier-induced bond softening, rather than a consequence of lattice anharmonicity~\cite{Murray2005}. Figure~\ref{figGabor} displays the continuous wavelet transform (CWT) of the transient reflectivity averaged in the energy range $24\pm0.5$~eV. The CWT used the Morlet wavelet as implemented in Matlab R2018a. In the first 0-1.5~ps, the chirp of the oscillations is clearly visible, with the instantaneous frequency evolving from $\approx 2.4$~THz to 2.9~THz. After~2.8 ps, the contrast of the oscillation becomes too low, meaning that the measured decrease can be attributed to noise.

\section{Computational methods}

\subsection{Numerical details}
\label{sec:Technical_details_of_calculations}

All calculations were performed using the \textsc{Quantum ESPRESSO} distribution~\cite{Giannozzi:2017, Giannozzi:2020} which is based on density-functional theory (DFT)~\cite{Hohenberg:1964, Kohn:1965}, plane-waves (PWs) basis sets, and pseudopotentials (PP). 

Bismuth crystallizes in the $A7$ rhombohedral structure, and the resulting lattice has the trigonal symmetry with two atoms in the primitive unit cell~\cite{Timrov:2012}. The rhombohedral cell is described by two parameters: the lattice parameter $a$, and the rhombohedral angle $\alpha$. The position of two Bi atoms in the unit cell is described by one parameter, the so-called internal parameter $u$ which describes the atomic positions in the crystallographic axes: one Bi atom has coordinates $(u,u,u)$ and the other one $(-u,-u,-u)$. We optimized these parameters using DFT with the setup described below and obtained $a=4.70$~\angstrom, $\alpha=58.00^\circ$, and $u=0.23559$, which is in very good agreement with the values obtained from x-ray diffraction measurements~\cite{Schiferl:1969}. 
\begin{figure*}[t]
 \includegraphics[width=.8\linewidth]{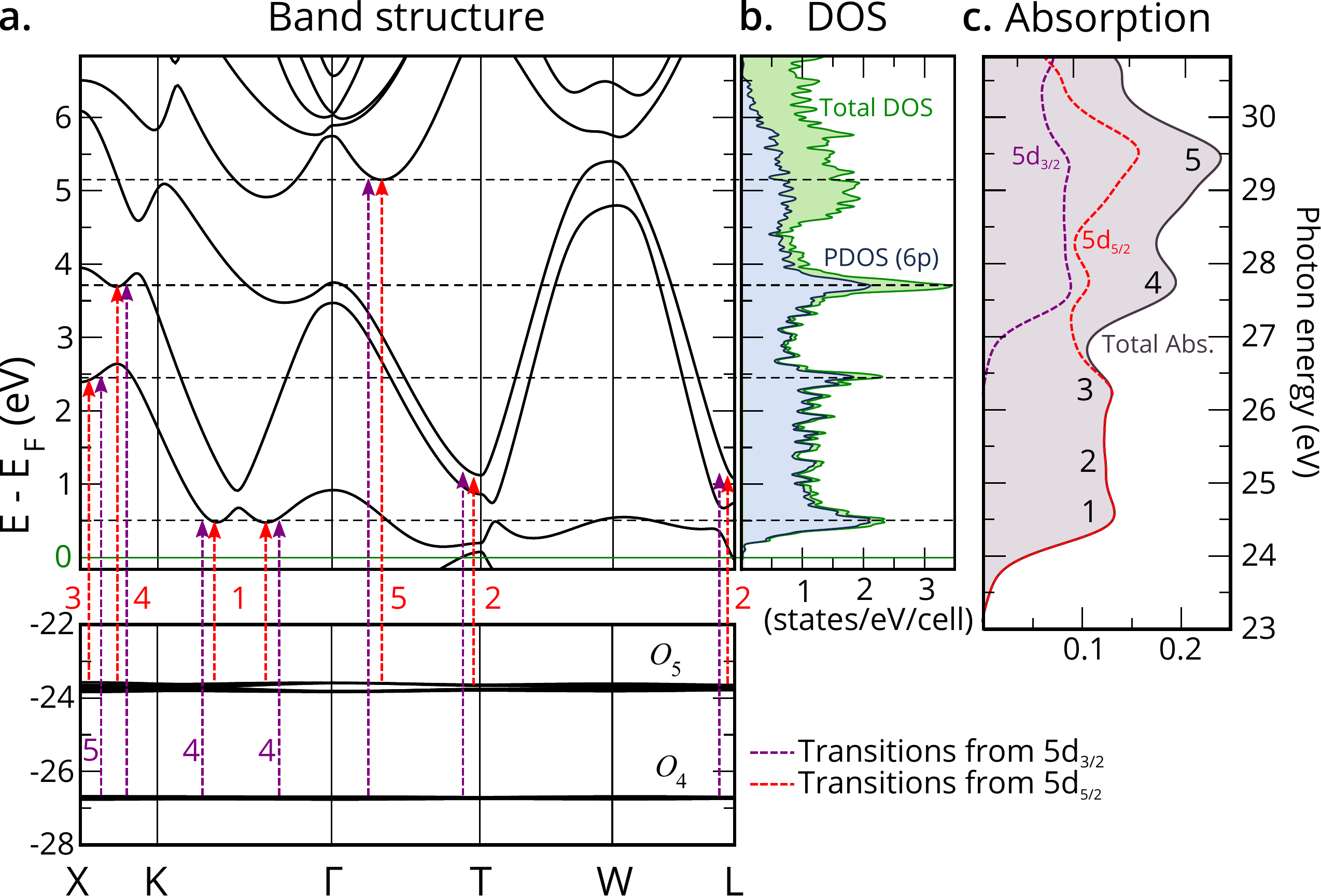}
 \caption{\label{figBands} Band structure and feature assignment. (a)~Band structure of bismuth computed using DFT. The binding energies of the $5d$ bands were rigidly increased by 2.7~eV. Vertical dashed arrows represent electronic excitations from the $O_4$ and $O_5$ shells to the lowest conduction states (only a few transitions are shown, but there are many others). The zero of energy is taken to be the Fermi level, $E_\mathrm{F}$, shown as a green horizontal line. (b)~Total density of states (DOS, green shaded area) and projected density of states (PDOS, blue shaded area) for $6p$ bands only (summed out for two atoms in the primitive unit cell). (c)~Absorption spectrum of bismuth as computed using TDDFT (shaded area), and decomposed into $5d_{3/2}$ and $5d_{5/2}$ (red and purple dashed line, respectively). The spectrum was rigidly shifted to higher energies by 2.7~eV.}
\end{figure*}

The ground-state properties of Bi (equilibrium structure and projected density of states) were computed using DFT with the following computational setup. We have used a fully-relativistic norm-conserving PP, by including $5d$, $6s$, and $6p$ electrons in the valence region, while keeping the remaining electrons frozen in the core region. The details about the PP can be found in Ref.~\cite{Timrov:2012}. The spin-orbit coupling effect was included in the calculations~\cite{Timrov:2012}, and for the xc energy we have used local density approximation (LDA) with the Perdew-Zunger parametrization of the Ceperley-Alder functional \cite{Perdew1981}. The Kohn-Sham wavefunctions were expanded in PWs with a kinetic-energy cutoff of 70~Ry, while the electronic charge-density and potentials were expanded in PWs with a cutoff of 280~Ry. To sample the Brillouin zone we have used a uniform $26 \times 26 \times 26$ $\mathbf{k}$ point mesh centered at the $\Gamma$ point, and we have used the Marzari-Vanderbilt smearing method~\cite{Marzari:1999} with a broadening parameter of $5 \times 10^{-3}$~Ry. The Fermi level was determined using a more refined computational setup as explained in Ref.~\cite{Timrov:2012}, namely by using a $50 \times 50 \times 50$ $\mathbf{k}$ point mesh and the tetrahedron method~\cite{Blochl1994}. This latter refinement is needed in order to describe accurately the electron and hole pockets in the vicinity of the Fermi level.

The excited-state properties of Bi (absorption spectrum) were computed using TDDFT~\cite{Runge:1984, Gross:1985}, with adiabatic local density approximation. The TDDFT equations were solved in the linear-response regime in the frequency domain using the recursive Liouville-Lanczos approach~\cite{Giannozzi:2017}. Local-field effects were taken into account. The absorption spectra were obtained by computing the macroscopic dielectric function $\varepsilon(\omega)$ which then gives access to imaginary part of the refractive index, i.e. $\mathrm{Im}[\tilde{n}](E)$, where $E$ is the excitation energy. Spectra have been convoluted with a Lorentzian function with a broadening parameter of $2 \times 10^{-2}$~Ry. We have computed absorption spectra for various distortions of the interatomic Bi-Bi distances to model the changes in the refractive index $\tilde{n}$ due to the coupling of carriers with the lattice vibrations after the perturbation by the external electric field (laser pulse). The changes in the Bi-Bi distance are expressed as $\delta = 100\% \times (d_\mathrm{new} - d_\mathrm{eq})/d_\mathrm{eq}$, where $d_\mathrm{eq}$ is the equilibrium distance between Bi atoms, and $d_\mathrm{new}$ is the new (increased or decreased) distance.

\subsection{Origin of peaks in the absorption spectrum}
\label{app:Origin_of_peaks}

Here we discuss briefly the origin of peaks in the absorption spectrum of bismuth based on the band structure and density of states (DOS). Figure~\ref{figBands}a shows the band structure of bismuth computed using DFT. The overall band structure is in very good agreement with previous studies~\cite{Timrov:2012, Faure2013}; in particular, the electron pockets at the L point and hole pockets at the T point are well reproduced. Due to the spin-orbit coupling, the $5d$ shell is split on two subshells, $O_5$ and $O_4$. The computed $5d$ bands were rigidly shifted to lower energies by 2.7~eV in order to better match (on average) the experimental binding energies of the $O_5$ and $O_4$ subshells relative to the Fermi level, which are at $-24.4 \pm 0.6$ and $-26.5 \pm 0.5$~eV, respectively~\cite{Bearden1967}. Figure~\ref{figBands}b shows the total DOS and projected DOS (PDOS) for the $6p$ bands in the energy range from 0 to 7~eV above the Fermi level. The peaks in the DOS and PDOS correspond to the extrema in the conduction bands manifold. 

In the photoexcitation process, electrons are promoted from the $O_5$ and $O_4$ subshells to the lowest conduction bands due to the absorption of photons of different energy. The resulting absorption spectrum is shown in Fig.~\ref{figBands}c, which was computed using TDDFT and split into $5d_{3/2}$ and $5d_{5/2}$ assuming a spin-orbit splitting of 3.1~eV and respective degeneracies of 4 and 6. It is instructive to discuss the origin of peaks in the absorption spectrum. Peaks labeled from 1 to 3 in Fig.~\ref{figBands}c are exclusively due to transitions from the $O_5$ subshell to the lowest conductions bands, while peaks 4 and 5 have contributions both from the excitations from the $O_5$ and $O_4$ subshells~\cite{Ejiri1981}. 

\subsection{Evaluation of the possibility of a phonon-induced topological phase transition in Bi}
\label{app:lpoint}

The energy separation between the two highest occupied electronic states at the L point in the Brillouin zone (which, for the sake of simplicity, we will refer to as the ``band gap at L'') is in the range from 11 to 15~meV according to experiments \cite{Maltz1970,Isaacson1969}. Such a small gap could be closed and even inverted, for instance by doping \cite{Jin2020}, thereby triggering a topological phase transition. Here we briefly explore whether a phonon-induced shift of the valence band could close the band gap at L.

From the computational side, reproducing such a small band gap very accurately is a significant challenge: DFT with local or semi-local xc functionals [such as LDA and generalized-gradient approximation (GGA)] largely overestimates the value of the band gap at L (see e.g. Table I in \cite{Aguilera2015}). More advanced computational approaches, such as quasiparticle self-consistent $GW$~\cite{Aguilera2015}, are needed in order to have accurate estimates of the band gap at L. Our current study is based on DFT-LDA (which is accurate enough to describe absorption properties of Bi in the energy range of tens of eV) and the band gap at L at equilibrium turns out to be 112 meV, which is larger than the experimental band gap at L by an order of magnitude, in accordance with \cite{Aguilera2015}. Hence, DFT-LDA is not an appropriate level of theory for the investigation of the intricate details of electronic states at the L point (and hence topological properties of Bi). 

\begin{figure}[h!]
 \includegraphics[width=.85\linewidth]{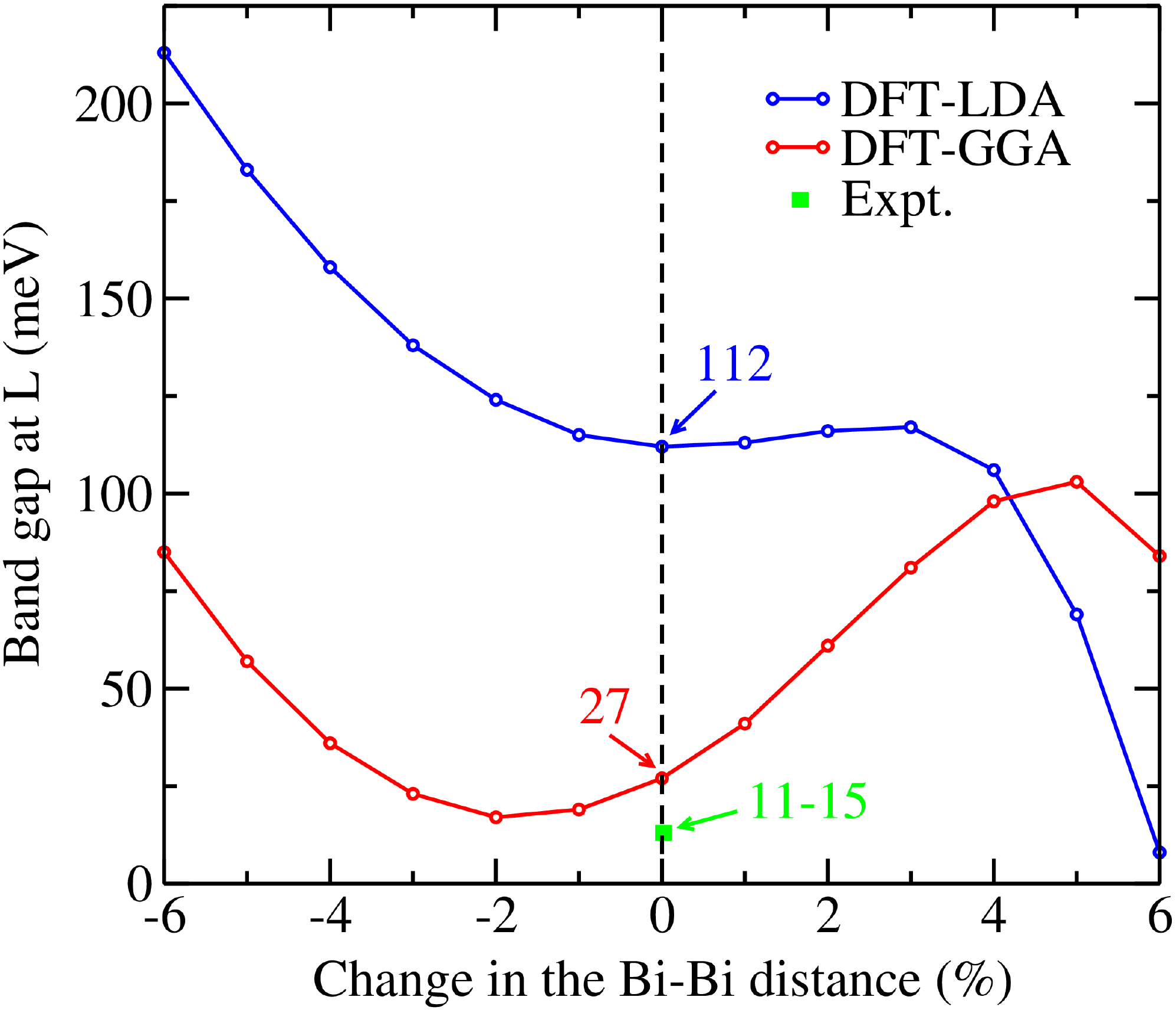}
 \caption{\label{fig_lpoint} Band gap at L (in meV) as a function of the change in Bi-Bi distance (in \%) computed using DFT with LDA and GGA xc functionals and as measured in experiments \cite{Maltz1970,Isaacson1969}.}
\end{figure}

Nonetheless, we computed the band gap at L as the distance between the two Bi atoms in the unit cell is changed, which corresponds to the A$_{1g}$ phonon mode excitation. In addition, we repeated this study using GGA with the Perdew-Burke-Ernzerhof (PBE) parametrization \cite{Perdew1996}, in order to compare the obtained trends; the results are shown in Fig.~\ref{fig_lpoint}. We can see that at the equilibrium Bi-Bi distance, GGA gives a band gap at L of 27~meV which is greatly improved with respect to LDA, but still it is a factor of $\sim 2$ larger than the experimental value. Interestingly, we see that increasing or decreasing the distance between Bi atoms gives different trends in LDA and GGA, but most importantly the band gap at L is not closed (for the magnitudes of Bi displacements considered here). This means that according to this level of theory, we cannot observe the gap closing and a subsequent re-opening, which would be a signature of the transition to a topologically insulating state \cite{Jin2020}. However, since the band gap at L is largely overestimated at LDA and GGA, these conclusions must be tempered. More advanced studies using DFT with xc functionals of higher accuracy (e.g. hybrid functionals \cite{Adamo1999,Heyd2003} or meta-GGA SCAN \cite{Sun2015}) or the quasiparticle self-consistent $GW$ method are needed in order to provide definitive conclusions on the effect of the A$_{1g}$ phonon on the band gap at L via the electron-phonon interaction.

\section{Decomposition of the transient reflectivity trace} 
\label{app:Decomposition}

At a fixed incidence angle $\theta$, the time-dependent reflectivity is linked to the complex index of refraction $\Tilde{n}$ through the Fresnel equation, here for s-polarized light:
\begin{equation*}
    R_s(E,t) = \left| \frac{\cos{\theta} - \Tilde{n}(E,t)\sqrt{1-(\frac{\sin{\theta}}{\Tilde{n}(E,t)})^2}}
    {\cos{\theta} - \Tilde{n}(E,t)\sqrt{1+(\frac{\sin{\theta}}{\Tilde{n}(E,t)})^2}}
    \right|^2 ,
\end{equation*}
where $\Tilde{n}$ reads:
\begin{align*}
    \Tilde{n}(E,t) = n(E,t) + \mathrm{i}k(E,t) .
\end{align*}

\begin{figure*}[t]
 \includegraphics[width=\linewidth]{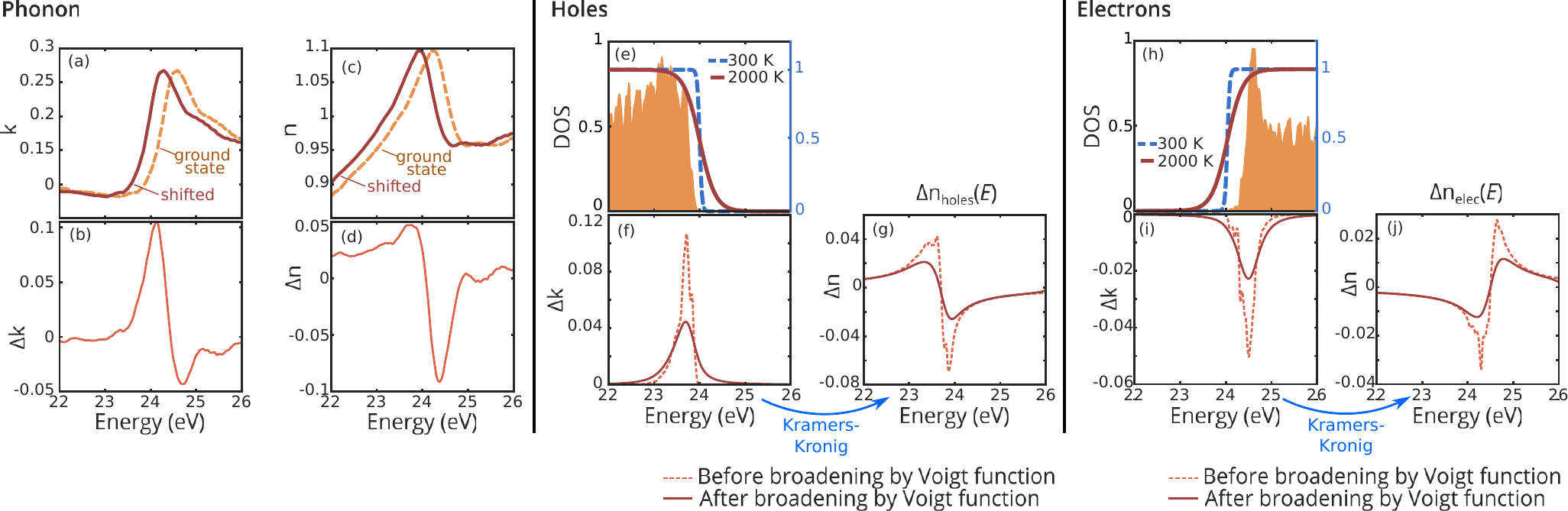}
 \caption{\label{figFitting} Contributions to the model transient reflectivity trace. For phonons, $k$ and $n$ are shifted (a and c), resulting in $\Delta k$ and $\Delta n$ shown in (b and d). For carriers, the change in $k$ is described as a change of temperature of the Fermi-Dirac distribution multiplied by the respective density of states, as explained in the text (e) and (h). The difference between the two yields $\Delta k$ for each carrier type (f) and (i), which is then converted to $\Delta n$ by Kramers-Kronig transformation (g) and (j). These contributions (dashed lines in (f), (g), (i) and (j)) are then broadened by a Voigt function to account for experimental and lifetime broadening (full lines in (f), (g), (i) and (j)). $n$, $\Delta n$, $k$ and $\Delta k$ are all unitless.} 
\end{figure*}

As explained in the manuscript, the phonon contribution is to produce an energy shift of the spectrum by an amount $\Delta E$. Figures~\ref{figFitting}a and c show the static values of $k(E)$ and $n(E)$, shifted here by 300 meV - a purposely large value to make changes more visible. Taking the difference between the curves yields the change in index (Figs.~\ref{figFitting}b and d):
\begin{align*}
    \Delta k_{phonon}(E,t) &= k(E) - k(E-\Delta E(t)),\\
    \Delta n_{phonon}(E,t) &= n(E) - n(E-\Delta E(t)).
\end{align*}

In order to describe carrier contributions, we begin by treating $k(E,t)$ because it allows to directly write space-filling effects~\cite{Zurch2017a}. Upon excitation, the temperature of both electrons and holes changes, which modifies the occupation of valence and conduction band states according to the Fermi-Dirac (FD) distribution, $f_\mathrm{FD}$. Figures~\ref{figFitting}e and~h display the FD function for 300 and 2000~K, on top of the density of states (DOS) for holes and electrons, respectively. The actually occupied states are obtained by multiplying the FD distribution with the DOS of valence and conduction bands ($N_{\mathrm{VB}}$ and $N_{\mathrm{CB}}$, respectively). The difference between the two thereby gives the states that become empty (resp. occupied) in the valence (resp. conduction) band. These give increased and reduced absorption, respectively (Figs.~\ref{figFitting}f and~\ref{figFitting}i). These quantities must then be broadened to account for both experimental and core-hole lifetime broadening ($\sigma$ and $\Gamma$, respectively), as shown by solid lines in Figs.~\ref{figFitting}f and ~\ref{figFitting}i. The changes in the imaginary part of the refractive index for electrons and holes thus read:
\begin{multline*}
    \Delta k_{elec}(E,t) = a_{elec} \left[- N_{\mathrm{CB}}(E) \biggl( f_{\mathrm{FD}}(E;T_{elec}(t)) \right. \\
    \left. - f_{\mathrm{FD}}(E;T_{0}) \biggr) \right]\otimes \mathcal{V}(E;\Gamma,\sigma),
\end{multline*}
\begin{multline*}
    \Delta k_{holes}(E,t) = a_{holes} \biggl[ N_{\mathrm{VB}}(E) \left(f_{\mathrm{FD}}(E;T_{holes}(t)) \right. \\
    \left. - f_{\mathrm{FD}}(E;T_{0}) \biggr) \right]\otimes \mathcal{V}(E;\Gamma,\sigma),
\end{multline*}
where $\mathcal{V}$ is the Voigt function, $a_{elec}$ and $a_{holes}$ are time-independent scaling factors which are needed because the amount of change in absorption as a function of the change of occupation at a given energy is unknown - this quantity depends on the oscillator strength of all semicore-to-conduction bands transitions across the Brillouin zone. We then use the Kramers-Kronig transformation with these quantities to obtain $\Delta n_{elec}(E,t)$ and $\Delta n_{holes}(E,t)$ (Figs.~\ref{figFitting}g and ~\ref{figFitting}j). Since $\Delta k_{elec}$ and $\Delta k_{holes}$ are non-zero only in a small energy range, common issues with performing the Kramiers-Kronig transformation (which requires evaluating integrals from 0 to $\infty$ with respect to $E$) are circumvented. 

\begin{figure}[t]
 \includegraphics[width=\linewidth]{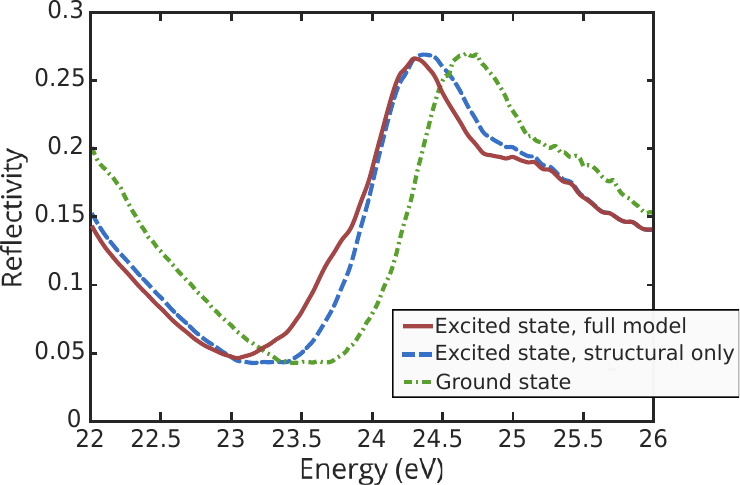}
 \caption{\label{figFittingResult} Modelled reflectivity using the contributions shown in Fig.~\ref{figFitting}, taking all contributions into account (full red line), only the phonons (dashed blue line) or no contributions (dash-dotted green line).}
\end{figure}

Finally, we sum the static value of $k$ with $\Delta k_{elec}$, $\Delta k_{holes}$ and $\Delta k_{phonon}$, and likewise for $n$. The obtained complex refractive index, $n(E,t)+\rmi k(E,t)$, is then used in the Fresnel equation to calculate the model reflectivity. Figure~\ref{figFittingResult} displays the result for this example case of a 300 meV shift and a carrier temperature of 2000 K for electrons and holes. Removing the carrier contribution highlights in which spectral region do they yield a reflectivity change. With this description in hand, the experimental transient reflectivity can be fitted by a least-square procedure. The parameters at each time delay are $\Delta E(t)$, $T_{holes}(t)$, and $T_{elec}(t)$, plus the two time-independent factors $a_{elec}$ and $a_{holes}$, resulting in a total of $N_t+2$ independent variables, where $N_t$ is the number of time points. We perform the least-square minimization on the 2D matrix (time $\times$ energy) using the parallel implementation of the \textit{lsqnonlin} routine in MATLAB 2018a. We used many randomly selected starting points as well as checked that lower and upper bounds did not influence the result. Finally, once the optimum was found, we computed the Jacobian at the solution point which was then used to obtain 95\% confidence intervals, shown together with the optimal solution in Fig.~\ref{fig4}. Lastly, we note that in principle there is a ``cross-term'' between phonon and carrier contributions: as the lattice displacement shifts the band positions, the DOS used in $\Delta k_{elec, holes}$ should be shifted as well. However, taking into account this effect adds two unknowns at each time delay (the VB and CB shifts), making the least-squares problem impossible to solve without large uncertainties for all parameters. Fortunately, this cross-term turns out to be a second-order effect: we verified that shifting the DOS in the hole and electron contribution by the maximal edge shift that we obtained (30~meV, see Fig.~\ref{fig4}) resulted in an error below 0.3\% in reflectivity. That is an order of magnitude below all contributions shown in Fig.~\ref{fig3}. Hence, we neglect this second-order effect to retain physically insightful quantities. The energy shift of the DOS arising due to this cross-term is most likely negligible in our case because of rather high carrier temperatures, and correspondingly wide Fermi-Dirac distributions, compounded with experimental and core-hole lifetime broadening. This might not be the case for narrower distributions.

\section{Comparison of measured atomic motion with literature}
\label{app:Atomic_motion}
\begin{figure}[h!]
 \includegraphics[width=\linewidth]{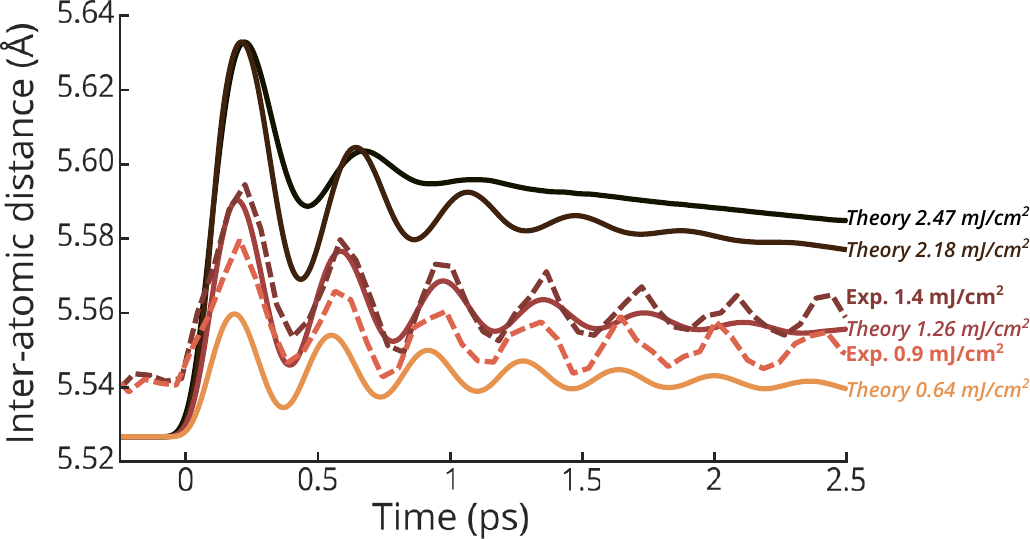}
 \caption{\label{figCompGiret} Comparison of measured displacement with computations from literature. Inter-atomic distance between the two Bi atoms in the primitive unit cell of bismuth as a function of time delay. Full lines are theoretical results from Ref.~\cite{Giret2011}, dashed lines are our experimental results (same as in Fig.~4 of the manuscript). The sample temperature before excitation is 0 and 300~K for theory and experiment, respectively.}
\end{figure}
Here we directly compare the atomic motion measured in our experiment (Fig.~\ref{fig4}) with existing data in the literature. In particular, Giret et al.~\cite{Giret2011} developed a thermodynamical model wherein most of the parameters were obtained from \textit{ab initio} calculations. The agreement between their result and experimental measurements of time-resolved x-ray diffraction~\cite{Fritz2007} is remarkable - not only in the general trend, but in the predicted magnitudes of motion. It is therefore a perfect benchmark to compare our results with.
Figure~\ref{figCompGiret} shows the inter-atomic Bi-Bi distance as a function of the pump-probe delay from Giret et al.~\cite{Giret2011} and from our measurements. The value before excitation is different because the initial temperature was 0 and 300~K in theory and experiment, respectively. While the calculations were not performed at the exact same pump fluences as our experiments, our data intercalates well between the theoretical curves. The agreement is striking: the chirped oscillations are well reproduced with consistent fluence-dependent frequencies, and the amplitude of displacement conforms to the ones predicted by theory. The only slight difference arises at very small time delays, which is to be expected for at least two reasons: (1)~the pump pulse durations are different (70~fs for theory compared to 5~fs in our case) and, (2)~as stated in the main manuscript, our description of carrier dynamics at small time delays is most likely flawed before they thermalize. This could impact the decomposition procedure and in turn influence the extracted atomic motion. In conclusion, this direct comparison suggests that broadband XUV transient reflectivity provides quantitative and accurate measurements of real-space atomic motion.

\bibliography{references_RG,references_IT}

\end{document}